\documentclass{PoS}
\usepackage{amsmath}
\usepackage{amssymb}
\usepackage{amsfonts}
\usepackage{subfigure}
\usepackage{soul}

\newcommand{\arxiv}[1]{arXiv:\,\href{http://arxiv.org/abs/#1}{{\tt #1}}}

\title{Charmonium resonances from 2+1 flavor CLS lattices}

\ShortTitle{Charmonium resonances from 2+1 flavor CLS lattices}

\author{Sara Collins$^{a}$, \speaker{Daniel Mohler}$^{c,d}$, M. Padmanath$^{a}$, \speaker{Stefano Piemonte}$^{a}$, Sasa Prelovsek$^{b}$, Simon Weish\"aupl$^{a}$ \\
\llap{$^{a}$} Institute for Theoretical Physics, University of Regensburg, 93040 Regensburg, Germany\\
\llap{$^{b}$} Faculty of Mathematics and Physics, University of Ljubljana, 1000 Ljubljana, Slovenia \\
\llap{$^{c}$} Helmholtz-Institut Mainz, 55099 Mainz, Germany\\
\llap{$^{d}$} Johannes Gutenberg Universit\"at Mainz, 55099 Mainz, Germany\\
        E-mail: \email{stefano.piemonte@ur.de}}

\abstract{Many exotic charmonium resonances have been identified recently in experiment, however their nature and properties are mostly unknown. Algorithmic and theoretical progress in lattice calculations has enabled reliable numerical investigation of the spectrum below the strong decay threshold, while the study of charmonium resonances remains an open challenge. The main difficulty to overcome is the presence of many open decay channels which are coupled together, resulting in a complex finite volume quantization condition. We report on our recent progress towards the determination of single-channel and coupled-channel scattering matrices in the scalar and vector channels on CLS ensembles. We also present an update concerning the study of the charmonium spectrum in moving frames.}

\FullConference{The 36th Annual International Symposium on Lattice Field Theory - LATTICE2018\\
		22-28 July, 2018\\
		Michigan State University, East Lansing, Michigan, USA.}
		
\begin{document}

\section{Introduction}

During the last decade collider experiments have been able to identify many exotic QCD resonances that do not fit in the conventional quark model. These are possible candidates for pentaquarks, tetraquarks or hadronic molecules. The discovery of new exotic states has led to an increasing interest in understanding their properties from a phenomenological point of view. Lattice Monte Carlo simulations can provide an {\it ab-initio} approach to the study of QCD resonances, however a full consideration of all aspects and systematics of resonances on the lattice is still an open challenge, especially in the heavy flavor sector.

Here we focus on the charmonium spectrum and we present the recent progress of our collaboration towards the study of excited charmonium states and exotic resonances. We first determine the charmonium spectrum  on the lattice in the single hadron approach, i.e. neglecting the effects of any strong decay modes. Here we extend the study of the spectrum from the rest frame \cite{Liu} to two moving frames \cite{PAD18}. We are also exploring the L\"uscher's finite volume method to determine scattering amplitudes and extract scalar and vector charmonium resonances, extending the first results of Ref.~\cite{Lang:2015sba}.

\section{Charmonium states in the single-hadron approach}

The starting point of any lattice hadron spectroscopy calculation is the determination of the two-point correlation functions $C_{ij}(t) = \langle O_i(t) O_j^\dag(0) \rangle$. The operators $O_i$ are classified according to their transformation properties on the lattice. Single- and multi- hadron operators are considered. The energy levels extracted from the solution of the Generalized EigenValue Problem (GEVP) using only single-hardon operators already provide information about the bound states of the charmonium spectrum and some indication about the resonant states above threshold. 

We have measured the charmonium correlation functions on the $N_f = 2+1$
ensembles labeled U101 and H105 generated by the CLS collaboration
\cite{Bruno:2014jqa,Bali:2016umi}. The lattice spacing is
$a=0.08636(98)(40)$~fm, and the pion and the kaon masses are equal to
$m_\pi\simeq 280$~MeV and $m_K\simeq467$~MeV. The lattice volume is $96\times 32^3$ for H105 and $128\times 24^3$ for U101. The gauge and the fermion fields fulfill open boundary conditions in the time direction. The bare charm quark mass $m_c$ is a parameter to be tuned. We have chosen two different values of $m_c$ corresponding to the $D$ meson mass being 80 MeV below and above its physical value to estimate the systematic errors and to explore how the charmonium spectrum and the exotic resonances are influenced by the mass of the charm quark.

We have first determined the charmonium spectrum in the single hadron approach on the ensemble U101 at rest and in moving frames. In both cases, many different lattice irreducible representations need to be considered in order to be able to identify the underlying continuum quantum numbers $J^P$ of the lattice energy levels. For further details about our results and our methodology, see Ref.~\cite{PAD18}.

\section{Scattering in a finite box: L\"uscher formalism}

The study of the physical properties of a resonance on a Euclidean space-time requires a different formalism that is beyond the simple estimation of energy
levels from the exponential decay of correlation functions of $\bar{c}c$-operators. The L\"uscher formalism relates the phase shift to the difference between the energy of the non-interacting system to the measured energy levels in a finite box \cite{Luscher}. The most challenging aspect of such a method is that a very precise determination of the lattice spectrum is required possibly on many different volumes or in many different momentum frames to reliably determine the scattering amplitudes.

In the elastic case, assuming that only a single scattering channel is coupled to the resonance under investigation, the L\"uscher formalism provides for each energy level a corresponding phase shift $p^{2l+1} \cot(\delta)$. In the case of two channels coupled together, the properties of the scattering matrix in the complex plane are accessible by fitting a suitable parameterization of the $K$-matrix to all energy levels \cite{MOR17}. We explore the elastic case in the vector channel and the coupled channel scenario for the scalar channel.

\subsection{Vector channel}

\begin{figure}[t]
  \subfigure{\includegraphics[width=.47\textwidth]{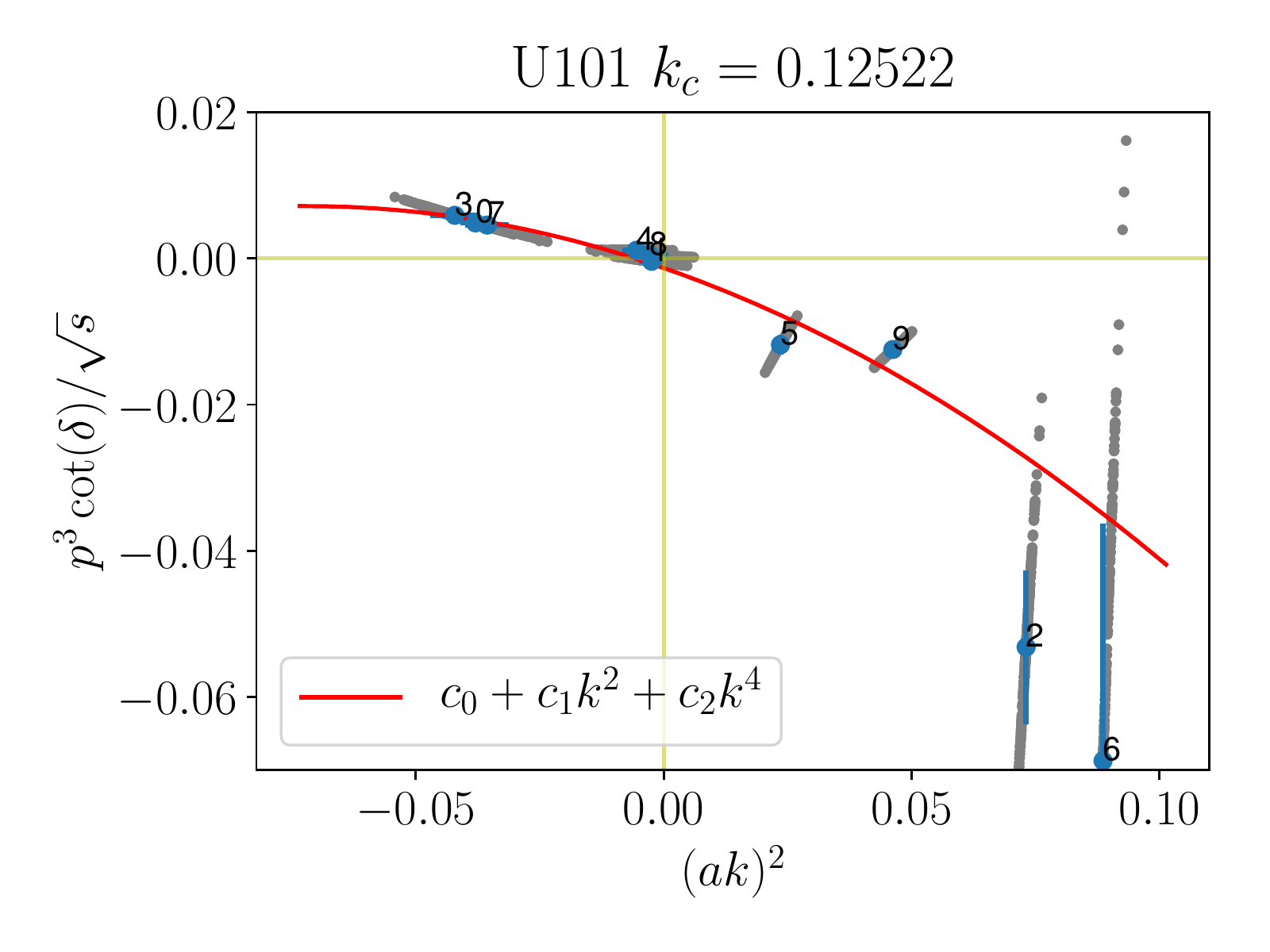}\label{U101_vector_channel}}
  \subfigure{\includegraphics[width=.47\textwidth]{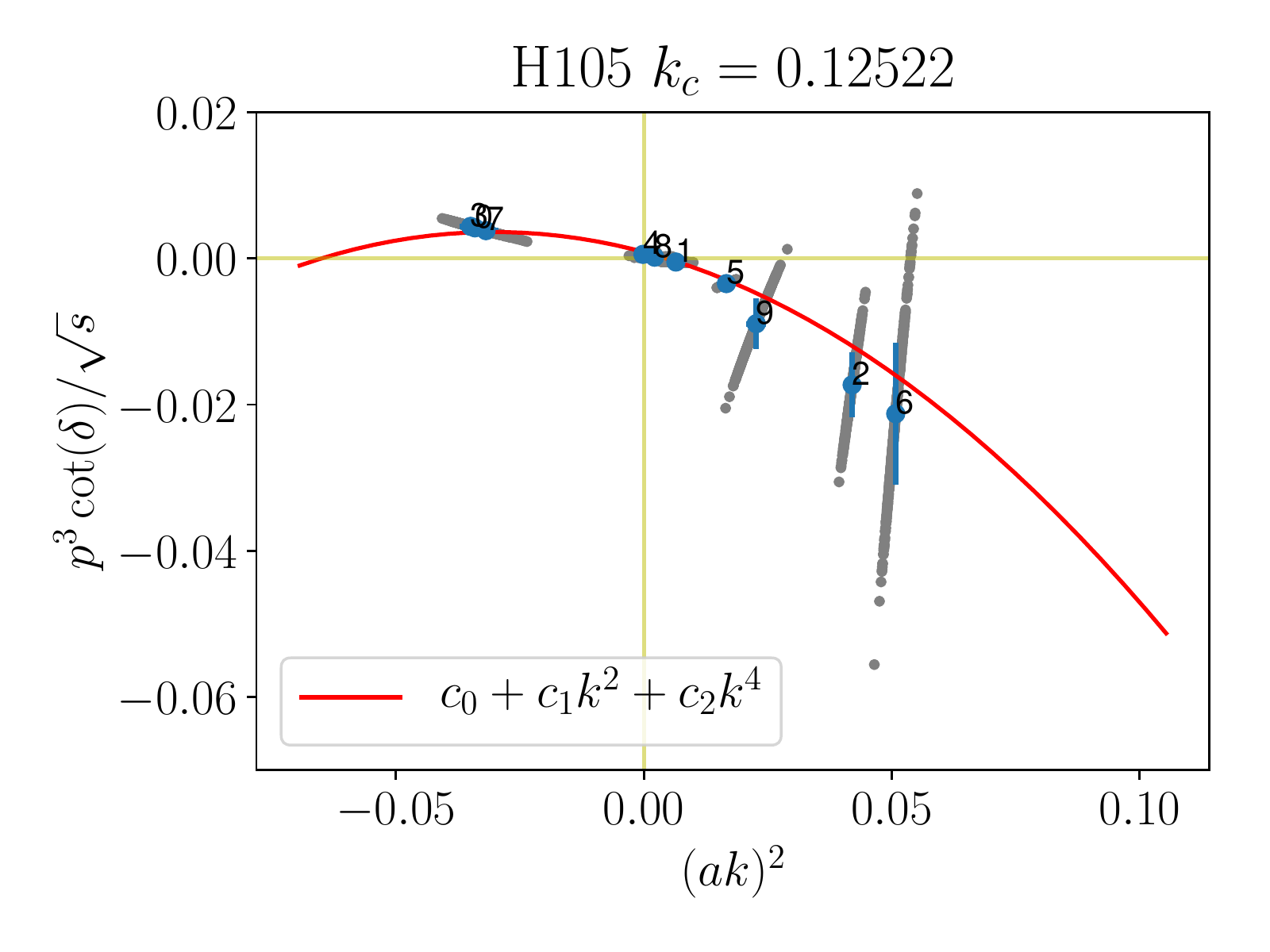}\label{H105_vector_channel}}
  \caption{Phase shift in vector channel as extracted from the energy levels at rest (points labeled 0-2) and in moving frames with momentum $P=(1,0,0)$ (points labeled 3-6) and $P=(1,1,0)$ (points labeled 7-9). The red curve represents a quadratic fit to our data.}
\end{figure}

The lowest states beyond the $J/\psi$ in the charm vector meson channel are $\Psi_{2S}$, $\psi(3770)$ and $\psi(4040)$. $\psi(3770)$ is the first resonance just above the open charm $\bar{D}D$ decay threshold. Therefore we include in our basis both at rest and in moving frames the corresponding two-particle $\bar{D}D$ interpolators, neglecting the effects from a $3^{--}$ resonance and charm annihilation diagrams.

The extracted phase shift as function of $k^2 = s/4 - m_D^2$ is presented for U101 in Fig.~\ref{U101_vector_channel} and for H105 in Fig.~\ref{H105_vector_channel} at the charm quark mass corresponding to $\kappa_c=0.12522$. The first natural assumption to interpolate our data and parametrize the phase shift is a quadratic fit of the form
\begin{equation}\label{param}
 \frac{k^3 \cot(\delta)}{\sqrt{s}} = A + B s + C s^2\,.
\end{equation}
We use the package TwoHadronsInBox for fitting our energy levels to the
L\"uscher formula \cite{MOR17}. The lowest energy level, corresponding to
$J/\psi$, is excluded from our fit. The quadratic \emph{ansatz} works
reasonably well for H105 with a $\chi^2/\textrm{d.o.f.} = 2.2$, while a higher $\chi^2/\textrm{d.o.f.}$
is obtained for U101, indicating that higher orders should possibly be included in Eq.~\ref{param}.

\begin{figure}
  \centering
  \subfigure[U101]{\includegraphics[width=.39\textwidth]{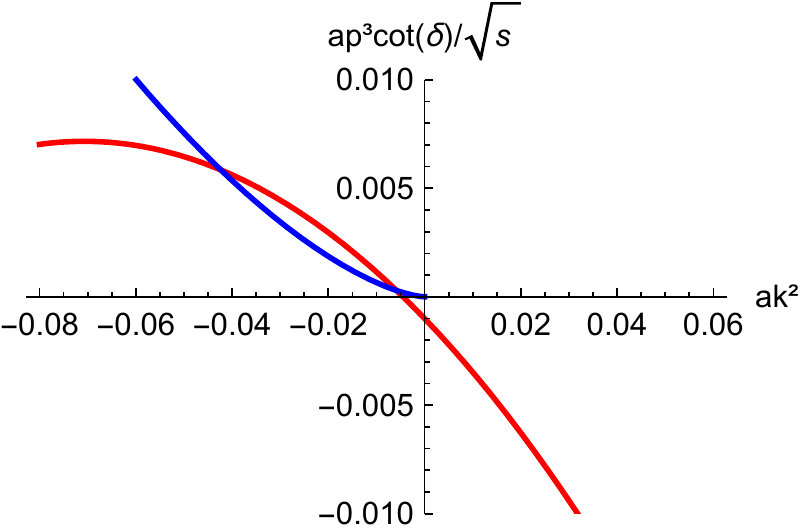}}
  \subfigure[H105]{\includegraphics[width=.39\textwidth]{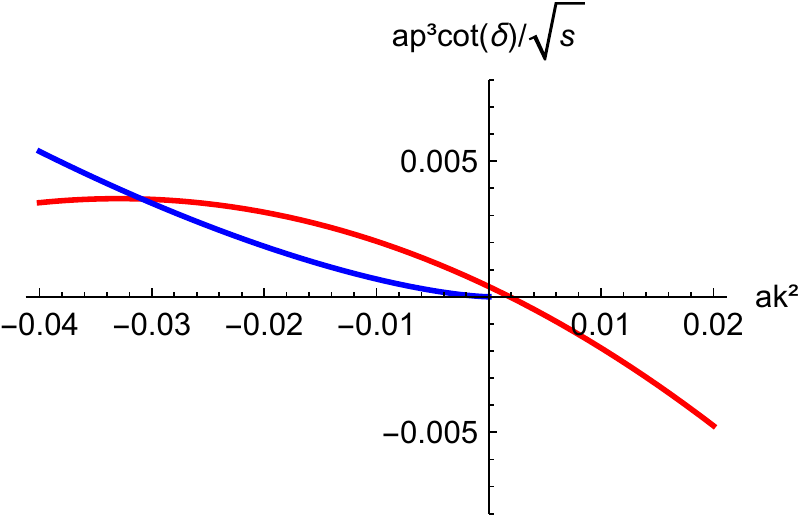}}
  \caption{Graphical representation of the pole condition for the presence of a bound state in $\bar{D}D$-scattering in the vector channel. The red and blue curves present $k^3 \cot(\delta)/\sqrt{s}$ and $- k^2 \sqrt{k^2}/\sqrt{s}$, respectively.} \label{pole_vector_channel}
\end{figure}

A bound state appears as a pole of the $T$ matrix on the real axis below the scattering threshold in the first Riemann sheet. Such a pole appears if the condition 
\begin{equation}
 k^3 \cot(\delta) = - k^2 \sqrt{k^2} 
\end{equation}
is fulfilled. The above condition is depicted in Fig.~\ref{pole_vector_channel}. For the ensemble U101 there are two crossings of the red and blue curves corresponding to two bound states, while for H105 there is a single crossing corresponding to the $\Psi(2S)$. Given that both ensembles have been generated with the same bare lattice parameters but different lattices sizes, our findings lead to the conclusion that the subleading exponentially suppressed finite volume effects
are relevant for U101 in this particularly delicate situation when  $\psi(3770)$ lies almost on threshold. It results from the fact that the points corresponding the naive energy levels for the $\psi(3770)$ lie at negative $k^2$ in Fig~\ref{U101_vector_channel}, so this result is independent of the parametrization chosen. Therefore we will focus on the H105 ensemble for the scattering analysis of the near-threshold resonance $\psi(3770)$.

\begin{figure}[h!]
  \subfigure[I]{\includegraphics[width=.47\textwidth]{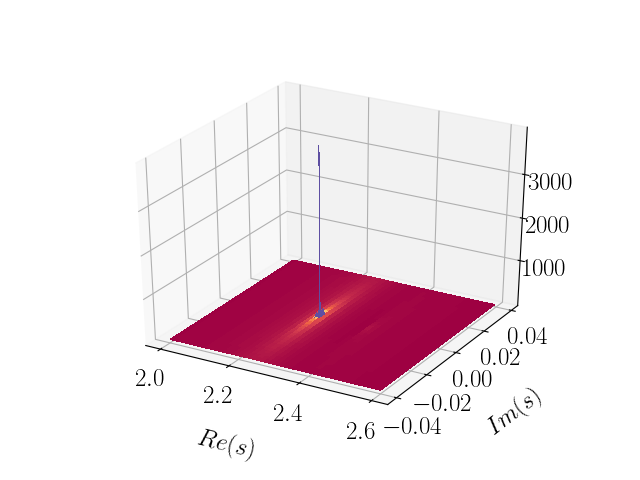}}
  \subfigure[II]{\includegraphics[width=.47\textwidth]{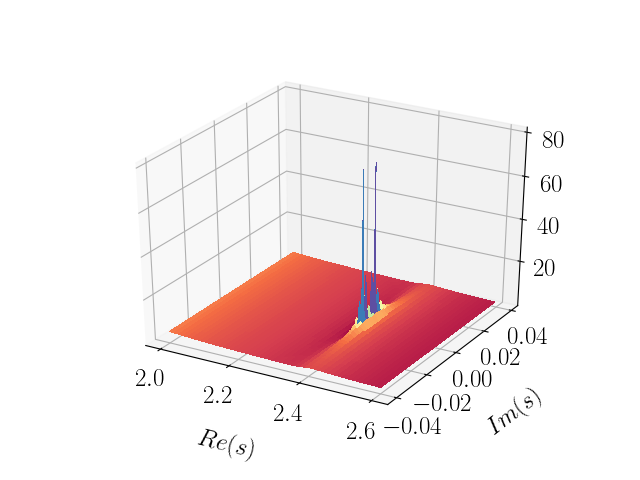}}
  \caption{The amplitude modulus for $\bar DD$ scattering in the vector channel plotted in the complex energy plane. The bound state ($\psi(2S)$) appears as a pole on the real axis of the first Riemann sheet, while the resonance ($\psi(3770)$) appears on the second Riemann sheet as poles off the real axis.}\label{complex_plane_vector_channel}
\end{figure}

The study of a scattering channel is performed by looking to the structure of the $T$ matrix in the complex $s$-plane. There is a single pole on the real axis of the $T$-matrix in the first Riemann sheet, corresponding to the bound state $\Psi(2S)$, and there are two complex conjugated pairs of poles in the second sheet corresponding to the $\psi(3770)$ resonance, see Fig.~\ref{complex_plane_vector_channel}. We have also explored different parametrization of our vector channel data to estimate the systematic errors and to find a functional form that is able to fulfill all physical requirements, such as the sanity check of Ref.~\cite{Aoki}. The full results of our analysis will be presented in a forthcoming publication.

\subsection{Scalar channel}

\begin{figure}[t]
\centering
\includegraphics[width=0.89\textwidth,height=0.31\textheight]{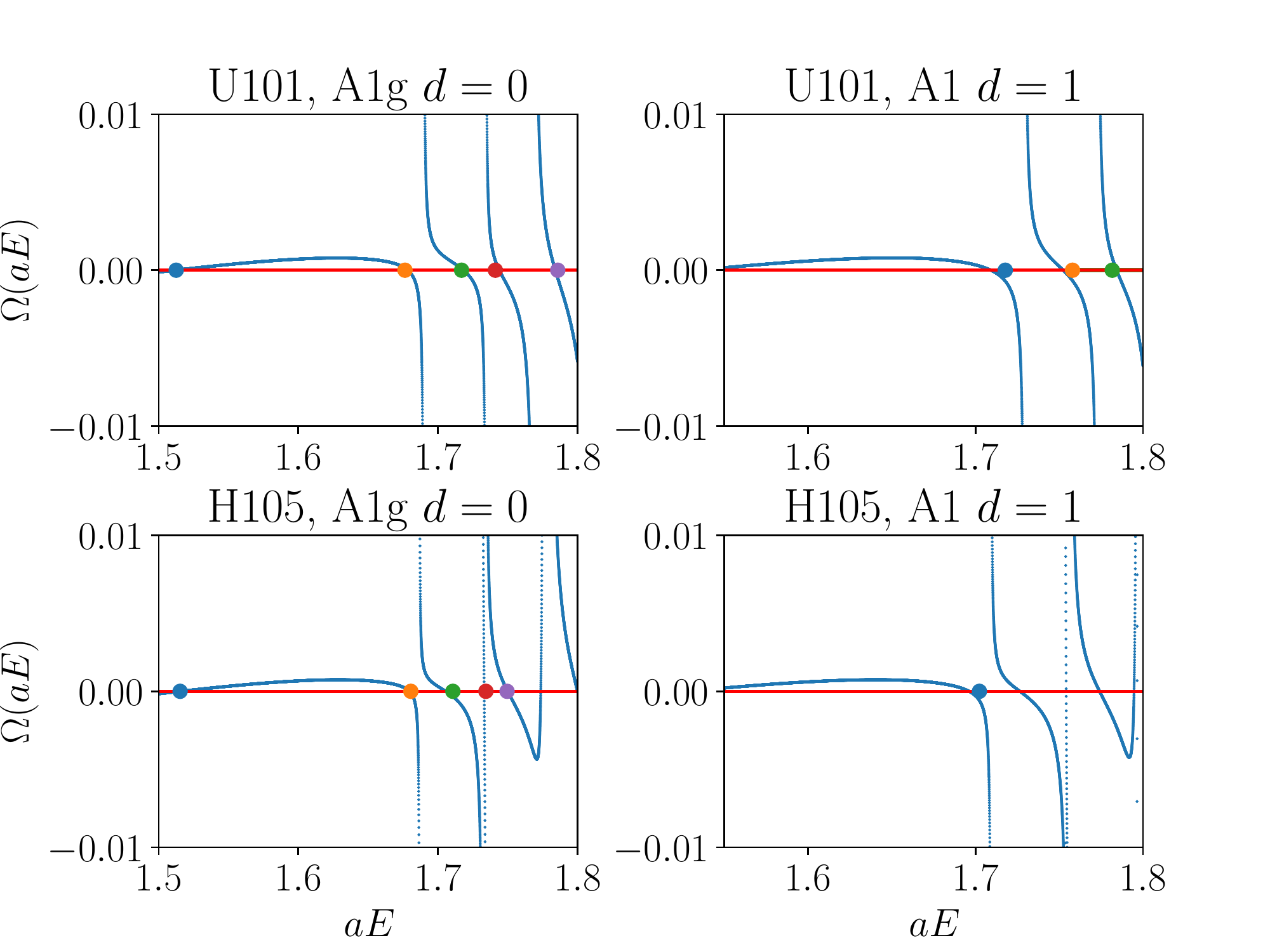}
\caption{$\Omega$ function for the scalar channel at $\kappa_c=0.12315$ and various lattice irreps. The current preliminary analysis does not include all energy levels on all irreps that have been measured.}\label{Omega_function}
\end{figure}

The $0^{++}$ charmonium channel is not yet experimentally well understood. Beyond the ground state, the $\chi_{c_0}$, two possible scalar resonances have been identified: the recently discovered $\chi_{c0}(3860)$ \cite{X3860} and the $X(3915)$ \cite{X3915}, for which an alternative interpretation as a $2^{++}$ state has been suggested in Ref.~\cite{Zhou:2015uva}. Both states are quite above the $\bar{D}D$ threshold and the $X(3915)$ is slightly below the $\bar{D}_s D_s$ threshold. In our setup we expect a significant influence of both $\bar{D}D$ and $\bar{D}_s D_s$ on our energy spectrum and we include their respective interpolators in our
basis for the GEVP. For this first analysis, we however exclude $J/\Psi-\omega$ and $\bar{D}^*D^*$ interpolators and we do not consider a possible $2^{++}$ resonance. We assume that these scattering channels have negligible influence.

If two decay channels are coupled, the $K^{-1}$-matrix can be parametrized in terms of three functions of $s$, the simplest \emph{ansatz} for them is a polynomial in powers of $s$. We have tried various possibilities and we find the form
\begin{equation*}
 \tilde{K}^{-1} = \begin{bmatrix}
    c_{11} s + b_{11} & b_{12} \\
    b_{12} & b_{22}
\end{bmatrix}
\end{equation*}
is the best compromise between constraining our fitting parameters while keeping a good $\chi^2/\textrm{d.o.f}\approx 1$ (notations for $\tilde{K}^{-1}$ are as in Ref.~\cite{MOR17}). In the coupled-channel case, the quality of fit can be judged by looking to the difference between our extracted energy levels and the zeros of the $\Omega$ functions in the various lattice irreducible representations. The $\Omega$ function, as introduced in Ref.~\cite{MOR17}, is defined as
\begin{equation}
\Omega(\mu,A(s)) = \frac{\det(A)}{\det((\mu^2 + AA^\dag)^{\frac{1}{2}})}\,,
\end{equation}
with $A(s)$ related to the box-quantization matrix and $\mu$ a regulator parameter. Our analysis is performed combining the U101 and H105 data from two different frames together. The quality of our fits is presented in Fig.~\ref{Omega_function}, where each point on the horizontal axis represents one of our lattice energy levels and the crossing of the blue curve to the real axis is the predicted energy level from our parametrization.

\begin{figure}
\centering
 \subfigure[I (+,+)]{\includegraphics[width=0.43\textwidth,height=0.17\textheight]{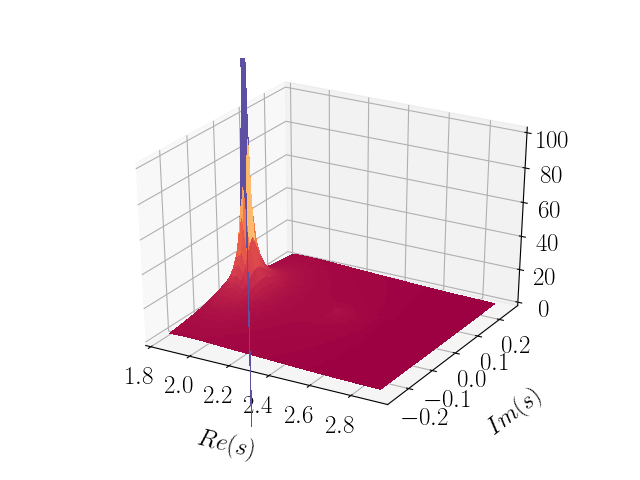}}
 \subfigure[II (+,-)]{\includegraphics[width=0.43\textwidth,height=0.17\textheight]{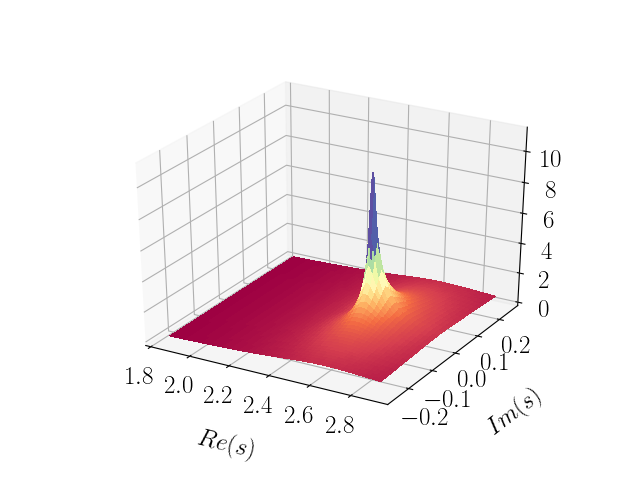}}
 \subfigure[III (-,+)]{\includegraphics[width=0.43\textwidth,height=0.17\textheight]{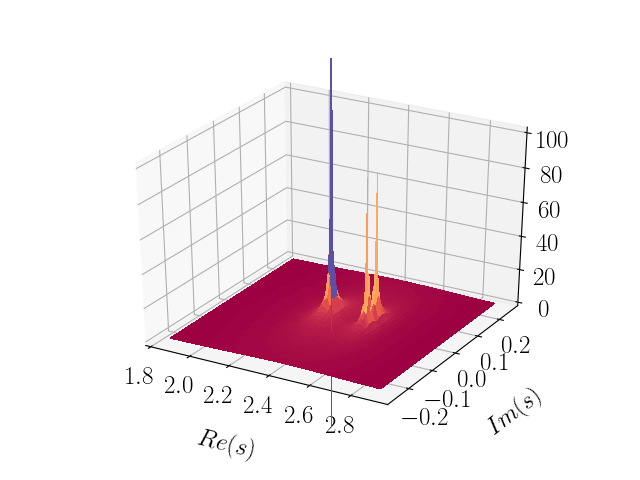}}
 \subfigure[IV (-,-)]{\includegraphics[width=0.43\textwidth,height=0.17\textheight]{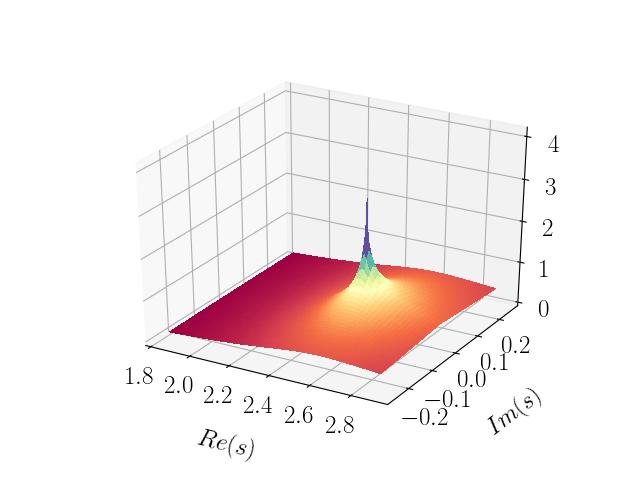}}
 \caption{$|T_{11}|$ plotted in the four complex energy Riemann sheets for $\bar{D}D$ and $D_s D_s$ scattering in the scalar channel. $|T_{11}|$ refers to the modulus of the scattering T-matrix element for the $\bar{D}D$ scattering channel. Note that the peaks in the Riemann sheets II and IV appear at $D_sD_s$ threshold and are related to the poles off the real axis in the Riemann sheet III.}\label{complex_riemann_zpp}
\end{figure}

In the case of two coupled channels, there are four Riemann sheets in the complex plane. These four sheets are labeled according to ($sign(Im[k_{m_D}]),~sign(Im[k_{m_{D_s}}])$), where $k_{m_i}^2 = s/4 - m_i^2$, in $\bar{D}D$ and $D_s D_s$ scattering. In Fig.~\ref{complex_riemann_zpp} we show the preliminary results for the $T$-matrix in the complex plane for each Riemann sheet. We plan to investigate the pole structures emerging in all complex planes and to extract the physical parameters of the scalar resonances in the near future.

\section{Conclusions}

The vector resonance $\psi(3770)$ requires a delicate tuning of the lattice volume and of the charm mass such that the strong decay in $\bar{D}D$ is allowed. The study of scalar resonances is challenging given the presence of multiple decay thresholds in the region of experimental interest, however we have been already able to identify a good parameterization of the $K$ matrix. There are still many systematic uncertainties and assumptions that have to be addressed to provide quantitative results that can be directly matched with experiments. 

\section{Acknowledgements}
We thank Ben H\"orz for his help and support with the package TwoHadronsInBox. Our code that implements the distillation method is written within the framework of the Chroma software package~\cite{Edwards:2004}. The inversion of the Dirac operator is performed used the multigrid solver of Ref.~\cite{Heybrock,Georg:2017diz}. The simulations were performed on the Regensburg iDataCool cluster, and the SFB/TRR 55 QPACE~2~\cite{Arts:2015jia} and QPACE~3 machines. We thank our colleagues in CLS for the joint effort in the generation of the gauge field ensembles  which form a basis for the here described computation. The Regensburg group was supported by the Deutsche Forschungs-gemeinschaft Grant No. SFB/TRR 55. M. P. acknowledges support from the EU under grant no. MSCA-IF-EF-ST-744659 (XQCDBaryons). S. Prelovsek was supported by Slovenian Research Agency ARRS (research core funding No. P1-0035 and No. J1-8137).

\end{document}